\documentclass[%
 reprint,
 amsmath,amssymb,
 aps,
]{revtex4-1}

\usepackage{graphicx}
\usepackage{graphics}
\usepackage{dcolumn}
\usepackage{bm}
\usepackage{float}
\usepackage{caption}
\usepackage{subfigure}
\usepackage{amsmath}
\usepackage{mathrsfs}
\usepackage{amssymb}

\begin{document}

\preprint{}

\title{Effective slip length of confined fluid with wavy fluid-solid interfaces}

\author{Li Wan}
\email{lwan@wzu.edu.cn}
\author{Yunmi Huang}
\affiliation{Department of Physics, Wenzhou University, Wenzhou 325035, P. R. China}
\author{Changcheng Huang}
\affiliation{Department of Computer Science, Wenzhou University, Wenzhou 325035, P. R. China}
\author{Rong An}
\affiliation{Department of Mathematics, Wenzhou University, Wenzhou 325035, P. R. China}
\date{\today}

\begin{abstract}
We have studied the effective slip length of confined fluid with nano-structured fluid-solid interfaces. A formula bridging the effective slip length and nano structures of the interfaces has been obtained analytically. For the application of the formula, a confined fluid with wavy fluid-solid interfaces has been studied as an example. In the study, the criteria of the formula have been detailed, and verified by the finite element method.  
\begin{description}
\item[PACS numbers]
47.10.ad, 47.15.-x, 47.61.-k, 87.85.Rs,47.63.mf
\item[Keywords]
effective slip length, confined fluid, nano-structured interface, wavy interface
\end{description}
\end{abstract}

\maketitle

\section{Introduction}
Nanofluidics with fluid confined in a scale down to a few nanometers exhibit many remarkable properties and have considerable potential in applications~\cite{Eijkel,Bocquet1,Karnik1,Karnik2,Schasfoort,Siwy}. In order to manipulate the transport behaviors of the fluid and develop new physical models of nano-scale effects of nanofluidics, the fluid-solid interfaces have been fabricated with various nano-structures by taking advantages of nowadays nano-technologies~\cite{Liu, Ou1,Ou2,Choi1,Choi2,Callies,Truesdell,Bocquet4,Joseph,Lauga,Richardson,Cottin4}. Studies on the nano-structured fluid-solid interfaces(NSFSI) have increased in the filed of nanofluidics, and rapid advances have been continuously made. For the various structures of the interfaces, an effective slip length has been introduced to evaluate the properties of the structures~\cite{Bocquet1}. That means, the nano-structured interfaces are replaced by flat interfaces and the effects of the nano structures are normalized into the effective slip length. The general expression for the effective slip length has not been obtained yet. In this work, we study this issue and figure out the expression analytically.\\

It is well known that the behaviors of fluid are governed by the Navier-Stokes (NS) equation, and it has been considered that the NS equation is still valid for the nanofluidics with the scale of a few nano meters ~\cite{Chan,Becker,Bocquet2,Georges,Leng,Li,Maali,Raviv}. To solve the NS equation, proper boundary conditions(BC) should be applied. The most commonly used BC is the no-slip BC, which requires the fluid velocity disappear at the fluid-solid interfaces. Such no-slip BC increases the hydrodynamic resistance to the fluid when the scale of the fluid is decreased ~\cite{Bocquet1,Bocquet2}. To overcome the drawback of the no-slip BC, a different BC called Navier BC has been introduced, and allows slippage of the fluid at the fluid-solid interfaces, meaning that the fluid velocity parallel to the interfaces could be nonzero ~\cite{Bocquet1,Bocquet2}. Various techniques have been developed by utilizing the Navier BC, such as by coting the fluid-solid interface with hydrophobic or hydrophilic materials~\cite{Ou1,Ou2}. \\

To identify the slippage of fluid at the interfaces, the slip length is introduced in the Navier BC ~\cite{Batchelor,Bocquet3,Pit,Cottin1,Cottin2,Thompson,Cieplak,Zhu,Priezjev,Jabbarzadeh}. The slip length is denoted by $b$ and defined as $b=\eta/\kappa$ with $\eta$ the viscosity of fluid and $\kappa$ the friction coefficient of the interface. Thus, the Navier BC takes the form of $\partial U/\partial l=-U/b$ with $U$ the fluid velocity component tangential to the interface and $l$ the displacement outward normal to the interface. And then the NS equation can be solved combined with the Navier BC. We note that $\eta$ close to the interface is enhanced compared to the viscosity in the bulk~\cite{Bocquet3,Hu,Israelachvili,Chan,Georges,Bocquet3,Chen}. However, the interfacial range for the enhanced $\eta$ takes only a few molecular layers, and does not change the structures of the interfaces. We can shift the fluid-solid interfaces by covering the interfacial range and absorb the effects of the interfacial range of enhanced $\eta$ in the intrinsic slip length $b_w$~\cite{Chen, Wan}. Thus, it is reasonable to set $\eta$ uniform in the whole computational domain. Generally, numerical methods such as the finite element methods should be used for the solving of the NS equation.\\

To explore physical properties of the slip length, nano-structured fluid-solid interfaces (NSFSI) have been developed for the nanofluidics. In order to catch the influence of the NSFSIs on the fluid dynamics, the finite element method to solve the NS equation has to mesh the fine structures of the interfaces heavily, especially for complicated structures. Such numerical solutions have computational consumption and the results are not universal for various NSFSIs. Actually, we can simplify the numerical solving by replacing the NSFSIs with flat interfaces, and normalize the physical effects of the NSFSIs into the effective slip length (ESL) for the flat interfaces. \\

We have studied the ESL for confined fluid with rough fluid-solid interfaces by statistically averaging the fluid velocity over the rough interfaces~\cite{Wan}. But for the nano-structured fluid-solid interfaces, the situation is different. Instead of the statistical averaging, we need figure out the ESL as a function of positions on the NSFSIs, which is exactly the main content of this work. To verify our result, we apply the ESL on a confined fluid with wavy fluid-solid interfaces and finite element method has been adopted for the verification.

\section{Theory}
We set two NSFSIs be normal to $z$ direction. The fluid confined by the two NSFSIs flows along $x$ direction. The system is uniform along $y$ direction for simplification. The position of points on the upper NSFSI is described by $z=h+\zeta_{u}(x)$ as a function of $x$, while the position of points on the lower one is by $z=-h+\zeta_{d}(x)$. The functions $\zeta_{u(d)}$ are averaged to be zero over the structured interfaces. Thus, the averaged locations of the two NSFSIs are at $z=h$ and $z=-h$ for the upper and lower interfaces respectively. In this study, we focus on only one interface, say the upper interface, since the other interface has the similar treatment. For convenience, we simplify the notation $\zeta_{u}$ by $\zeta$ in the following. The intrinsic slip length of the NSFSI is denoted by $b_w$, which is understood as an intrinsic parameter of the system originated from the chemical interaction between the fluid molecules and the solid molecules at the interfaces. The ESL $b$ involves the informations of $b_w$ and the nano structures of the interfaces.

\subsection{General result}
For the nanofludics with the scale down to a few nano meters, the Reynolds number is comparable to or even smaller than 1. Thus, for the system in this study, the complete NS equation is reduced to the Stokes equation, reading
\begin{equation}
\label{Stokesequ}
\nabla \times \nabla \times \vec{U}=\frac{1}{\eta}\vec{f},
\end{equation} 
by ignoring the convect effect of the fluid and using the condition of incompressible flow~\cite{Wan}. Here, $\vec{U}$ is the fluid velocity and $\vec{f}$ represents forces including the pressure gradient and body forces in the fluid. Now we introduce a dyadic Green function $\overset{\leftrightarrow}{G}(\vec{r}, \vec{r}')$, satisfying the equation $
\nabla \times \nabla \times \overset{\leftrightarrow}{G}=\overset{\leftrightarrow}{I}\delta(\vec{r}-\vec{r}')$ with the same BC applied for eq.(\ref{Stokesequ}). Note that $\overset{\leftrightarrow}{I}$ is the dyadic unit and $\vec{r}(\vec{r}')$ is the position vector in the fluid. By combining the two equations and using the Green theorem, we solve out the fluid velocity as
\begin{align}
\label{solutionU}
\vec{U}(\vec{r}')=\int \hspace{-0.3cm} \int K ds +\int \hspace{-0.3cm} \int \hspace{-0.3cm} \int \frac{1}{\eta}\vec{f}(\vec{r})\cdot \overset{\leftrightarrow}{G}(\vec{r},\vec{r}')dV .
\end{align}
with $K=- (\hat{n} \times \nabla \times \vec{U})\cdot \overset{\leftrightarrow}{G}- (\hat{n} \times \vec{U})\cdot (\nabla \times \overset{\leftrightarrow}{G})$ ~\cite{Wan}. On the right hand side of eq.(\ref{solutionU}), the first term of $\iint K ds$ is for the integral over the total interface of the computational domain, while the second term is for the volume integral. In the expression of $K$, $\hat{n}$ is the unit vector outward normal to the interfaces. Note that the direction of the $\hat{n}$ varies on the interfaces due to the nano structures.\\

As we have mentioned, the fluid velocity in the real system with the NSFSIs can be obtained from an effective system in which the NSFSIs are replaced by flat interfaces and the effects of the NSFSIs are normalized in the ESL $b$ for the flat interfaces. In the effective system, $b$ now is a function of $x$. We assume that the amplitudes of the nano-structures are small enough that the fluid velocity at NSFSIs can be expanded around the flat interfaces in perturbation. Then the surface integral in eq.(\ref{solutionU}) can be transformed to a surface integral over the flat interfaces of the effective system with the ESL $b$ involved. The equivalence of the two fluid velocities in the real system and the effective system requires that $b$ should satisfy the following equation
\begin{align}
\label{bequ}
\frac{N}{b_w}(1-\frac{\zeta}{b})^2+\frac{\partial ^2 \zeta}{\partial x^2}N^{-2}(1-\frac{\zeta}{b})^2=\frac{1}{b}
\end{align}
with $N=\sqrt{1+(\frac{\partial \zeta}{\partial x})^2}$. This equation is different to the equation eq.(27) in Ref(\cite{Wan}) by the point that we do not take the statistical average over the interfaces and leave $b$ as a function of $x$. In the derivation of eq.(\ref{bequ}), the Navier BC has been applied. Details for the derivation can be found in the Ref(\cite{Wan}). Now $b$ can be solved out from eq.(\ref{bequ}) to be
\begin{align}
\label{ESL}
b=\frac{1+2\zeta A+\sqrt{1+4\zeta A}}{2A},
\end{align}
with $A=\frac{N}{b_w}+\frac{1}{N^2}\frac{\partial^2 \zeta}{\partial x^2}$ for short notation. It means that if we have the configuration $\zeta(x)$ of the NSFSIs, we can get the ESL $b$ directly from eq.(\ref{ESL}). Then we can perform the numerical calculation in the effective system for the fluid  velocity by using the ESL $b$. The fluid velocity obtained in the effective system is expected to be equivalent to the velocity in the real system if the ESL $b$ is valid.\\

For the applications we have to be careful that eq.(\ref{ESL}) is valid only if the condition $1+4\zeta A \ge 0$ is hold. What is more, eq.(\ref{ESL}) is obtained based on the perturbation expansion. Therefore, the condition of $1+4\zeta A \ge 0$  and the validity condition of the perturbation expansion both should be satisfied before the application of eq.(\ref{ESL}). These two conditions are dependent on the detail structures of the NSFSIs. As an example, we study the criteria of eq.(\ref{ESL}) on a real system with wavy interfaces.

\subsection{Wavy interfaces}
We consider that the wavy interfaces are symmetric and take $\zeta=M\cos(2\pi x/L)$ for the upper interface and $\zeta=-M\cos(2\pi x/L)$ for the lower one. Here, $L$ is the period of the wavy structure and $M$ is the amplitude. $M$ should be small enough for the validity of the perturbation expansion and also satisfies the condition of $1+4 \zeta A\ge 0$. The validity condition for the perturbation expansion has no explicit expression, which will be studied by the numerical calculation later. In this subsection, we only focus on the condition of $1+4 \zeta A\ge 0$.\\

We substitute $\zeta=M\cos (2\pi x/L)$ into the term A for the upper interface. After some algebra, the condition of $1+4\zeta A\ge 0$ is simplified to be
\begin{align}
\frac{(1+\xi ^2)b_w}{6M \xi^2}\ge G(g)
\end{align}
with $\xi=2\pi M/L$, $g=\cos (2\pi x/L)$ and 
\begin{align}
G(g)=g(g^2 +\frac{5b_wg}{6M}-\frac{2+3\xi^2}{3\xi^2}).
\end{align}
Now we try to find the maximum value of $G(g)$. It is known that the equation $G(g)=0$ has three roots denoted by $g_1$, $g_2$ and $g_3$. We arrange the roots in the sequence of $g_1<g_2<g_3$. Obviously, we have $g_1<0$, $g_2=0$ and $g_3>0$. And $G(g)$ is positive only when $g$ is in the ranges of $g_1<g<0$ and $g>g_3$. The solution $g_0$ maximizing $G$ in the range of $g_1<g<0$ can be obtained from the equation $G'(g)=\frac{\partial G(g)}{\partial g}=0$, which reads
\begin{align}
g_0=-\frac{5b_w}{18M}- \frac{1}{6}\sqrt{\frac{25b_w^2}{9M^2}+\frac{8+12\xi^2}{\xi^2}}.
\end{align}
But if it is $g_0<-1$, the maximum value for $G$ is $G(-1)$ instead of $G(g_0)$ in the range of $g_1<g<0$ due to $g=\cos (2\pi x/L)$. In the range of $g>g_3$, the possible maximum value for $G$ is $G(1)$. Combing all the informations above, we have the following criteria
\begin{subequations}
\label{criteria}
\begin{equation}
\frac{(1+\xi ^2)b_w}{6M \xi^2}\ge \max [G(1), G(-1)],~~~if~~~g_0<-1
\end{equation}
\begin{equation}
\frac{(1+\xi ^2)b_w}{6M \xi^2}\ge \max [G(1), G(g_0)],~~~if~~~g_0>-1
\end{equation}
\end{subequations}
for the validity of eq.(\ref{ESL}). Here, we take $\max [a,b]=a$ if $a>b$. Otherwise, we take $\max [a,b]=b$. The criteria eq.(\ref{criteria}) is independent on the variable $x$. The maximum amplitude $M_{max}$ can be obtained from the criteria if the structure parameters $L$, $b_w$ are given.

\subsection{Numerical calculation}
To verify our theory, we use FreeFem++ for the numerical calculation by taking the weak form of the Stokes equation~\cite{Hecht}. In the weak form of the equation, the viscosity $\eta$ has been absorbed in the pressure since we are only interested in the streamlines and velocities of the fluid. The numerical calculations are performed for the comparison of the two systems. One system is the real system with the NSFSIs and the other system is the effective system with flat interfaces and the ESL.\\

Fluxes for the two systems are kept the same. For each system, the fluid velocity is periodic at the inlet and the outlet to remove the side effect. We can achieve the periodicity by the following steps. At the first step, we free the outlet and obtain a velocity distribution at the outlet with a given input at the inlet. At the second step, we use the velocity distribution at the outlet as the input at the inlet. We repeat the two steps to get the periodicity of the fluid velocity at the inlet and the outlet finally. On the fluid-solid interfaces, the fluid velocity is governed by the Navier BC along the tangential direction and fixed to be zero along the normal direction due to the fact that fluid can not penetrate the solid interfaces. In the calculations, all the lengths of the systems are scaled by the parameter $h$.

\section{Results}
Fig. 1 shows an example of streamlines for the two systems. 
\begin{figure} 
\centering 
\includegraphics[width=4cm]{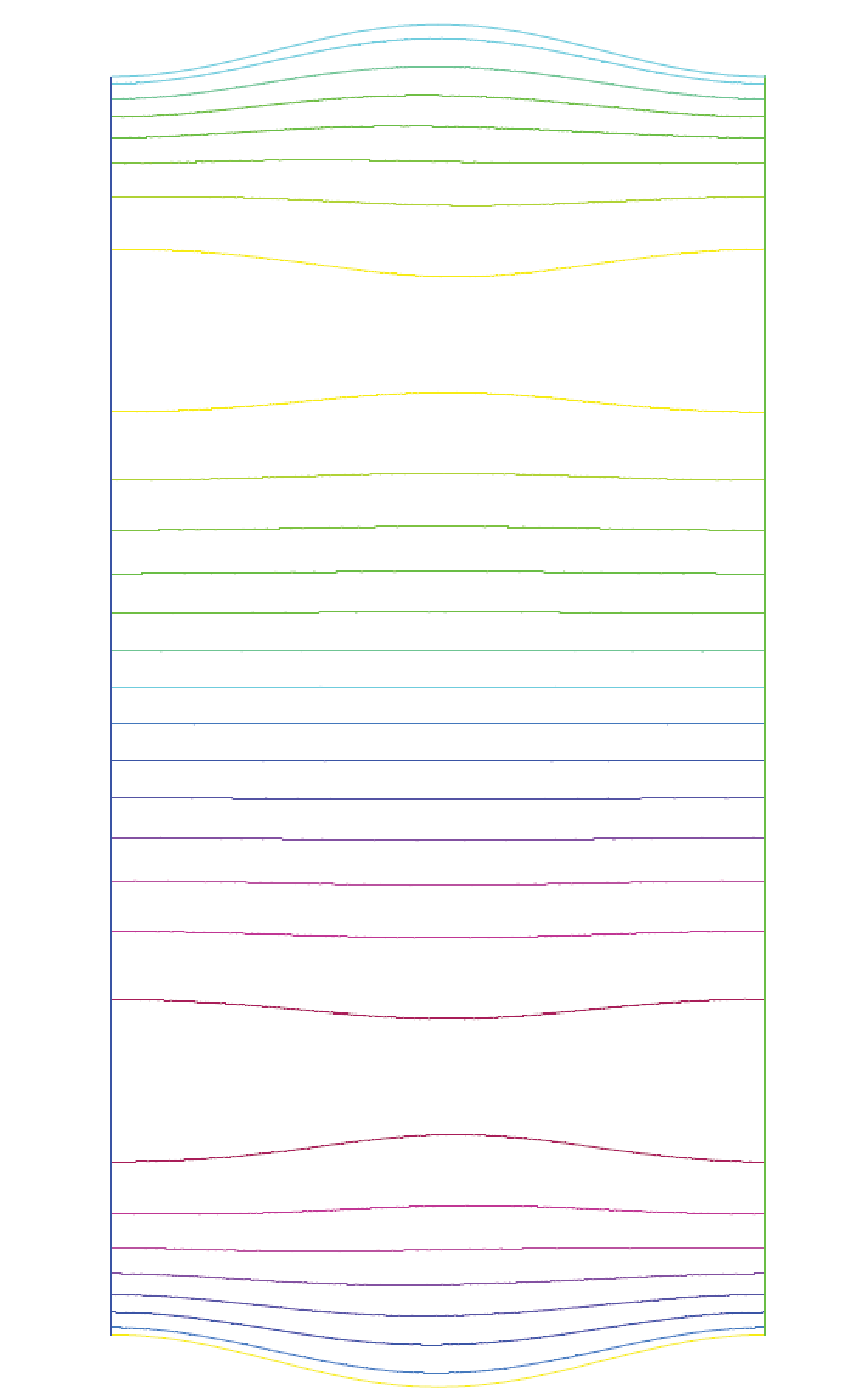} 
\includegraphics[width=4cm]{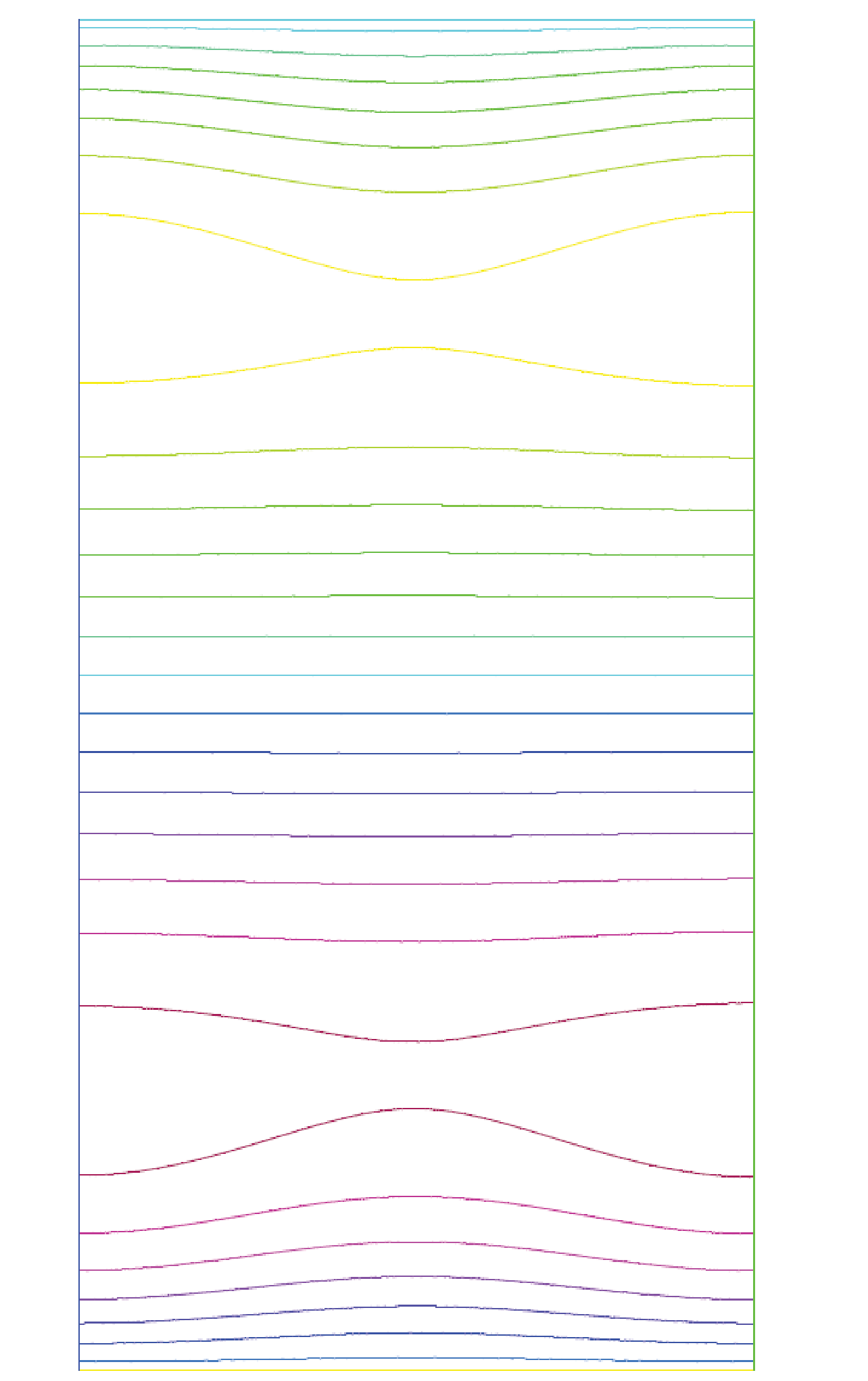} 
\caption{Streamlines for the two systems. The left figure is for the real system with the wavy interfaces. The right one is for the effective system with flat interfaces and ESL involved. Here, $L=1$, $M=0.04$ and $b_w=0.5$ have been used for the calculation with the input flux equaling to 2. The scaling $h=1$ has been applied. }   
\end{figure}
The left system in fig.1 is the real system with the wavy interfaces while the right one is the effective system with flat interfaces and ESL involved. As we have mentioned, the lengths in the systems have been scaled by $h$. Thus, in fig.1, the distance between the two plates is fixed to be 2. We take $L=1$, $M=0.04$ and $b_w=0.5$ for the calculation with the input flux equaling 2. The streamlines in the figure are symmetric with respect to $z$ axis. It seems from the figure that the streamlines in the effective system catches the main feature of those in the real system especially in the range away from the interfaces, showing the validity of our effective slip length.\\

In this work, we do not intend to study the streamlines, but focus on the deviation of the physical properties of the two systems. Since we consider the transport behaviors of the fluid, we study the deviation of the $x$-component velocities for the two systems. We define a quantity in the following by
\begin{align}
\label{devia}
\epsilon=\frac{\int |u_1-u_2| ds}{\int u_1 ds}.
\end{align}
Here, $u_1$ is the $x$-component fluid velocity of the real system while $u_2$ is of the effective system. The area integral in eq.(\ref{devia}) is over the computational domain limited in the range of $[-h+M, h-M]$ along $z$ direction.\\

We plot $\epsilon$ as a function of $M$ in fig.2. 
\begin{figure} 
\centering 
\includegraphics[width=8cm]{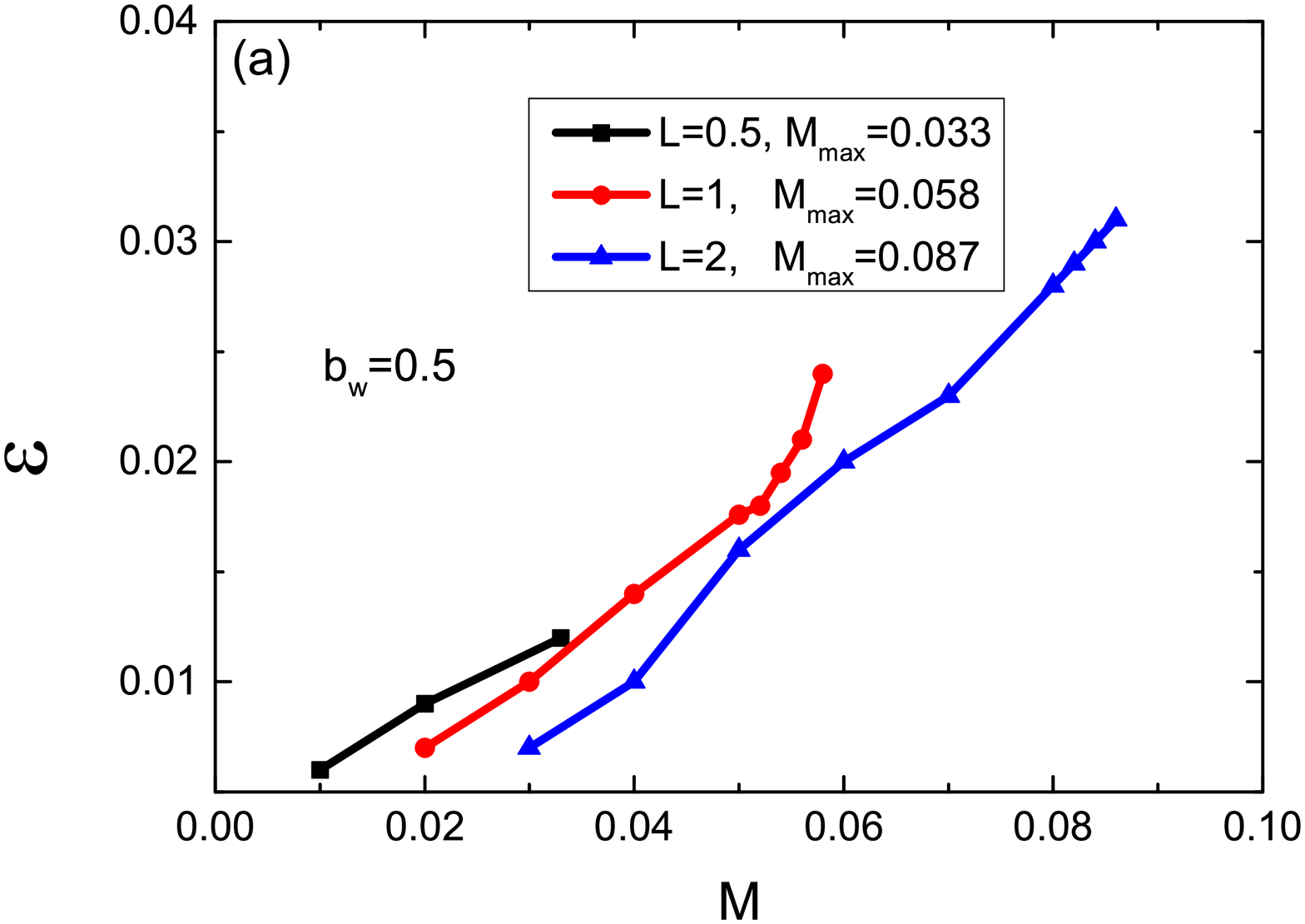} 
\includegraphics[width=8cm]{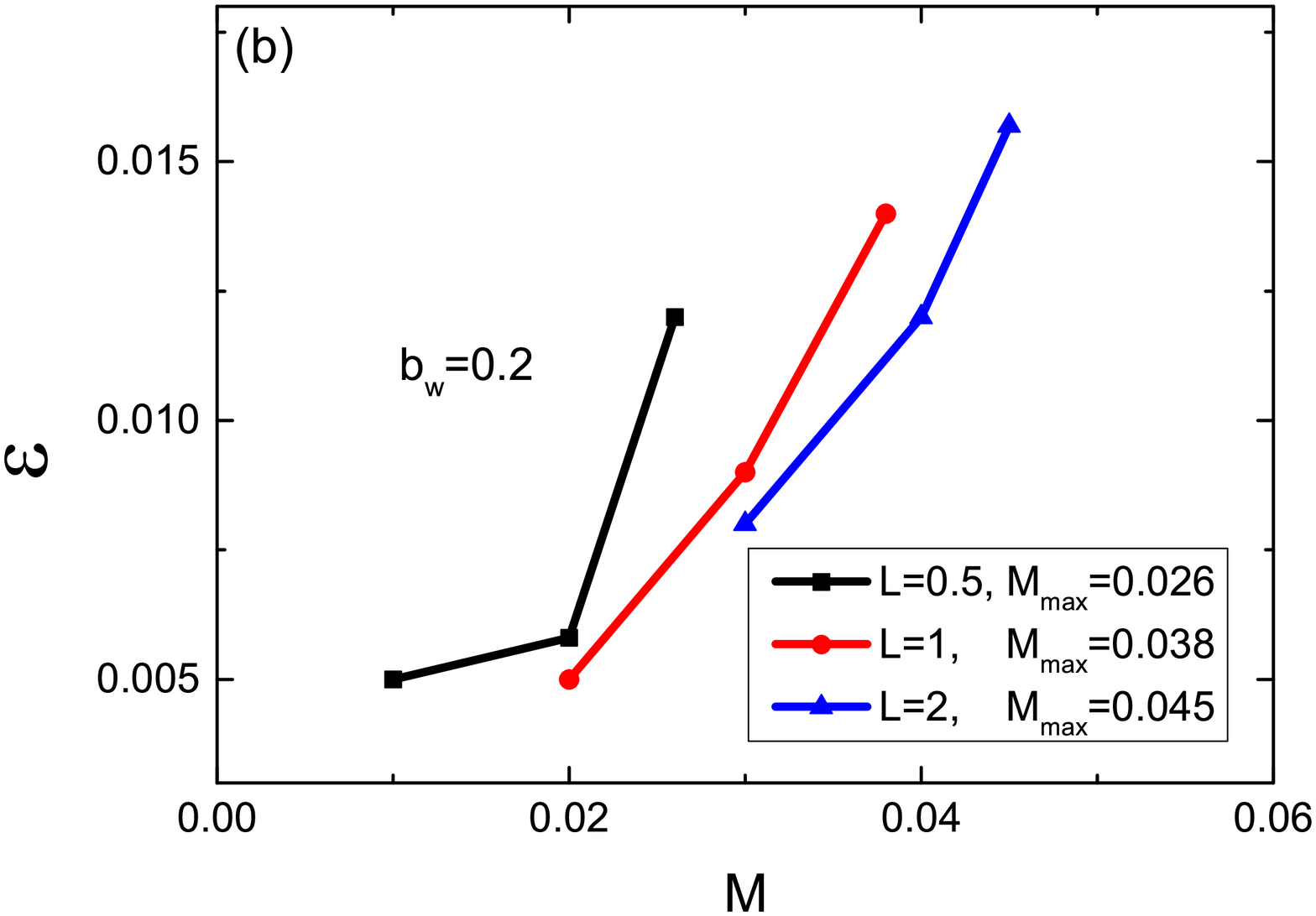} 
\caption{Deviation of $x$-component fluid velocities of the two systems. }   
\end{figure}
It shows that the deviation of the effective system from the real system, denoted by $\epsilon$, increases with the increasing of the amplitude  $M$ of the wavy interfaces. For $M=0$, the two systems coincide. The quantity $M_{max}$ indicated in the figure is solved from eq.(\ref{criteria}) and is the maximum of $M$ to guarantee the existence of the ESL. That means, the ESL is a real number instead of a complex one only when $M\le M_{max}$ is hold. Such condition has been checked by the numerical calculations, in which the ESL has no solution when $M$ exceeds $M_{max}$. It is expected that $\epsilon$ should decrease with the increasing of the period $L$ since the increasing of $L$ weakens the effect of $M$. Such expectation has been confirmed in fig.2 for a fixed $M$. \\

We have observed from the calculations that for the cases with $\epsilon >0.02$, the streamlines in the real system distort from the symmetric pattern and are asymmetric with respect to $z$ axis. The distortion is exactly the contribution to the deviation $\epsilon$. To decrease the deviation, we need involve the distortion in the ESL, such as by introducing the perturbation expansion not only along $z$ axis but also along $x$ direction at the effective flat interfaces in eq.(23) of ref(\cite{Wan}). Unfortunately, no proper BC like the Navier BC to treat such expansion term along $x$ direction, which remains a problem in our next work and is not in the scope of this study. For the application, we suggest that the perturbation expansion is acceptable for the cases of $\epsilon<0.02$ and then the ESL is valid in these cases. In fig.2, we also observe that the fluid with a larger intrinsic slip length $b_w$ brings a larger $\epsilon$ by comparing fig.2(a) and (b). Such phenomena is originated from the fact that the large slip length enhances the distortion of the streamlines close to the interfaces, which dominates the deviation. As we have suggested, such distortion can be ignored for the cases of $\epsilon<0.02$.

\section{Conclusions}
We have studied the effective slip length for confined liquid with nano-structured fluid-solid interfaces. The analytical expression of the effective slip length has been obtained. For the demonstration, we have applied the effective slip length on a system with wavy interfaces. The criteria of the expression have also been analyzed, which is restricted by two conditions. One condition is that the amplitude of the wavy interfaces should not be too large for the existence of the effective slip length, shown as $1+4\zeta A\ge 0$. The other condition is that the period of the wavy interfaces and the intrinsic slip length of the system should not be too large to break the perturbation expansion. No explicit expression is for this condition. We suggest that the perturbation condition is acceptable when the deviation $\epsilon <0.02 $ is hold.


\end{document}